\documentclass[a4paper]{jpconf}
\usepackage{graphicx}
\usepackage{amssymb}
\pdfoutput=1
\begin{document}

\title{The PHENIX Decadal Plan: Crafting the Future of RHIC}

\author{Christine A. Aidala for the PHENIX Collaboration}

\address{Los Alamos National Laboratory, Los Alamos, NM 87545}

\ead{caidala@bnl.gov}

\begin{abstract}
After ten years of running, the PHENIX Collaboration is starting to lay out the compelling physics that RHIC, as the most versatile hadron collider in the world, will be well positioned to explore beyond the program planned for upgrades already in progress.  As the fields that RHIC set out to investigate have advanced and evolved, new questions and directions have arisen at the frontiers of QCD, and we have exciting opportunities before us to continue to confront the challenges and surprises of strong interactions into the next decade.  The latest thoughts of the PHENIX Collaboration on this ongoing planning process for the future of RHIC are discussed.
\end{abstract}

\section{Studying QCD at RHIC}

The Relativistic Heavy Ion Collider (RHIC), as the most versatile hadron collider in the world, has provided an outstanding facility for the study of quantum chromodynamics (QCD) since it began taking data in 2000.  Its independent rings allow collisions of a variety of symmetric and asymmetric nuclear species, and past RHIC runs have collided deuterons on gold, copper on copper, and gold on gold.  As the world's only polarized proton collider, RHIC has also achieved polarized proton-proton collisions over energies ranging from $\sqrt{s}=62.4$~GeV to 500~GeV.

These different colliding systems allow the study of a variety of QCD matter.  The proton is the simplest stable bound state in QCD; investigating cold nuclear matter reveals how nuclei cannot merely be described as groups of protons and neutrons but rather the roles that collective effects play; the QCD matter formed by central heavy ion collisions, strongly interacting quark-gluon plasma (sQGP), is a deconfined state of QCD matter, too hot for hadronic bound states to exist.  Just as many areas of experimental condensed matter research seek to isolate or create a physical system and study its properties, such is the nature of the various programs at RHIC.  RHIC was built with a maximum collision energy for heavy ions comfortably above the energy needed to create deconfined QCD matter, and with great flexibility to explore a range of energies below this.  Given the complexity of nuclear collisions and QCD systems in general, the RHIC facility was designed from the beginning with $p+p$, $d+A$, and $A+A$ collisions at the same center-of-mass energy per nucleon all as integral parts of the program, enabling a comprehensive program in QCD physics and permitting more complex systems to be understood in terms of simpler ones.

After the tremendous success of the first decade of running RHIC, the community is now looking toward the next ten years.  A Decadal Planning document, which represents only a first step in the ongoing process of planning for the future of the facility, was submitted by the PHENIX Collaboration to BNL Management in September 2010~\cite{BNLnpp,PHENIXdp}.  As the several fields under study at RHIC have advanced and matured over the past decade, new physics questions have come to the forefront as we start to move into more quantitative investigation of QCD physics.

\noindent Questions relevant to the hot, dense QCD matter program at RHIC include:
\begin{itemize}
  \item Are quarks strongly coupled to the QGP at all
distance scales?
  \item What are the detailed mechanisms for parton-QGP
interactions and responses?
  \item Are there quasiparticles at any scale?
  \item Is there a relevant screening length in the QGP?
  \item How is rapid equilibration achieved?
\end{itemize}

\noindent The study of QCD processes in cold nuclear matter (CNM) centers on the following fundamental questions:
\begin{itemize}
\item What are the dynamics of partons at very small and very large
  momentum fraction ($x$) in nuclei, and at high gluon density what
  are the nonlinear evolution effects (i.e., saturation)?
\item What are the pQCD mechanisms that cause energy loss of partons
    in CNM, and is this intimately related to transverse momentum
    broadening?
\end{itemize}

\noindent Questions to be addressed mainly within the proton-proton program at RHIC include:
\begin{itemize}
  \item What is the 3D spatial structure of the nucleon?
  \item What is the nature of the spin of the nucleon?
  \item What spin-momentum correlations exist within hadrons and in the process of hadronization?
  \item What is the role of color interactions in different processes?
\end{itemize}

\begin{figure}[h]
\includegraphics[width=0.9\linewidth]{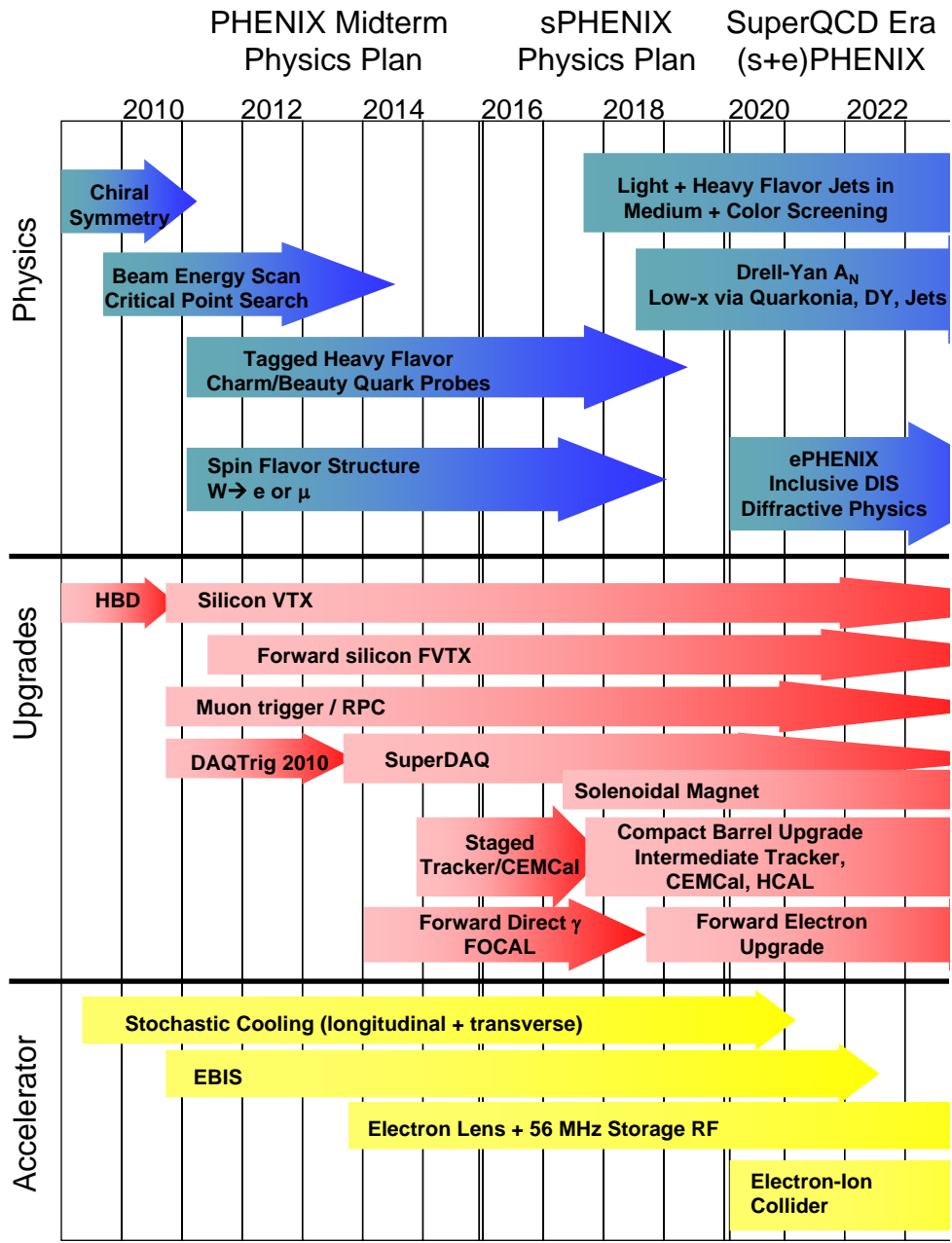}%
\caption{\label{fig:timeline}Approximate timeline indicating the physics topics, detector upgrades, and accelerator upgrades at RHIC over the next decade, taken from~\cite{PHENIXdp}.}
\end{figure}

Improved and extended machine and detector capabilities will enable us to more fully address these questions over the course of the next decade.  An approximate timeline indicating the physics topics, detector upgrades, and accelerator upgrades at RHIC over the next decade is shown in Fig.~\ref{fig:timeline}.

\section{An Even More Flexible Machine for the Future}

Quantitative understanding will develop from having a variety of measurements to compare, and upgrades in place or under development for RHIC will enable an even wider range of systems to be studied.

The electron beam ion source (EBIS) upgrade, already available as of 2011, has replaced the Tandem Van de Graaff accelerators as the heavy ion pre-injector for RHIC and will provide even greater flexibility in the choice of nuclear species to accelerate, regardless of their atomic electronegativity.  Uranium-uranium collisions, which will create an even higher energy density than gold-gold collisions, are planned for the near future as a first step in exploring these new capabilities.

Research and development are underway to study the prospects for polarized beams of helium-3 at RHIC, which would serve as effective polarized neutron beams, complementing the polarized proton beams already available.

Discussions are ongoing regarding the possibility of replacing the RHIC DX magnets to achieve energies up to $\sqrt{s}=650$~GeV for $p+p$ collisions, or $\sqrt{s}=260$~GeV per nucleon for Au+Au collisions.  While increasing the center-of-mass energy somewhat could be useful in order to increase the production rate of certain probes (e.g. $W$ bosons, heavy flavor), the parton momentum fraction ($x$) range of interest to the nucleon structure program is $x \sim$0.0001--0.5, and many intriguing effects seem to lie in the region $x \gtrapprox 0.01$, so greatly increased center-of-mass energies would be undesirable.  In the case of the heavy ion program, moderately higher energies would again increase the production rate of certain probes, but early data from the Large Hadron Collider (LHC) indicate that even at the significantly higher center-of-mass energies of 2.76~TeV per nucleon achieved for lead-lead collisions there, the deconfined QCD matter created is qualitatively the same as at RHIC.

\subsection{Adding an Electron Ring to RHIC}
The field of heavy ion physics already takes advantage of final-state probes that do not interact strongly, such as direct photons and internal conversions of thermal photons, as well as $Z$ bosons at the LHC.  By adding an electron ring to RHIC, in the future it will be possible to select electroweak or colored probes not only in the \emph{final} state but also the \emph{initial} state.  The availability of an electron and possibly a positron beam at RHIC will open up a wealth of new possibilities.  Measuring the energy and angle of the scattered electron allows one to reconstruct the kinematics of the scattered parton.  A lepton beam also provides a very clean probe of the initial hadronic or nuclear state, without the complications of strong interactions.

\begin{figure}[ht]
  \centering
  \raisebox{0.15in}{\includegraphics[width=0.54\linewidth]{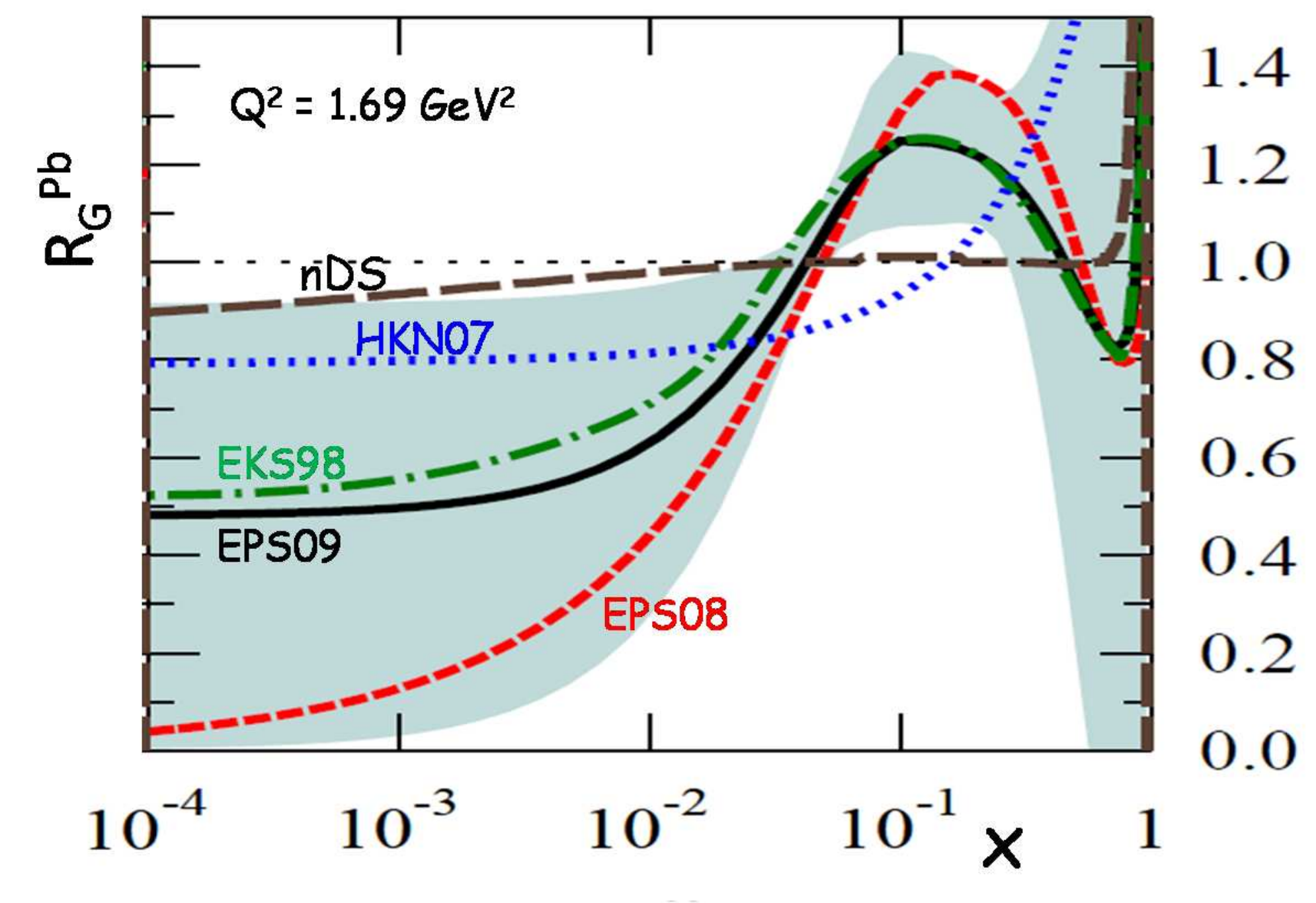}}
  \hfill
  \includegraphics[width=0.45\linewidth]{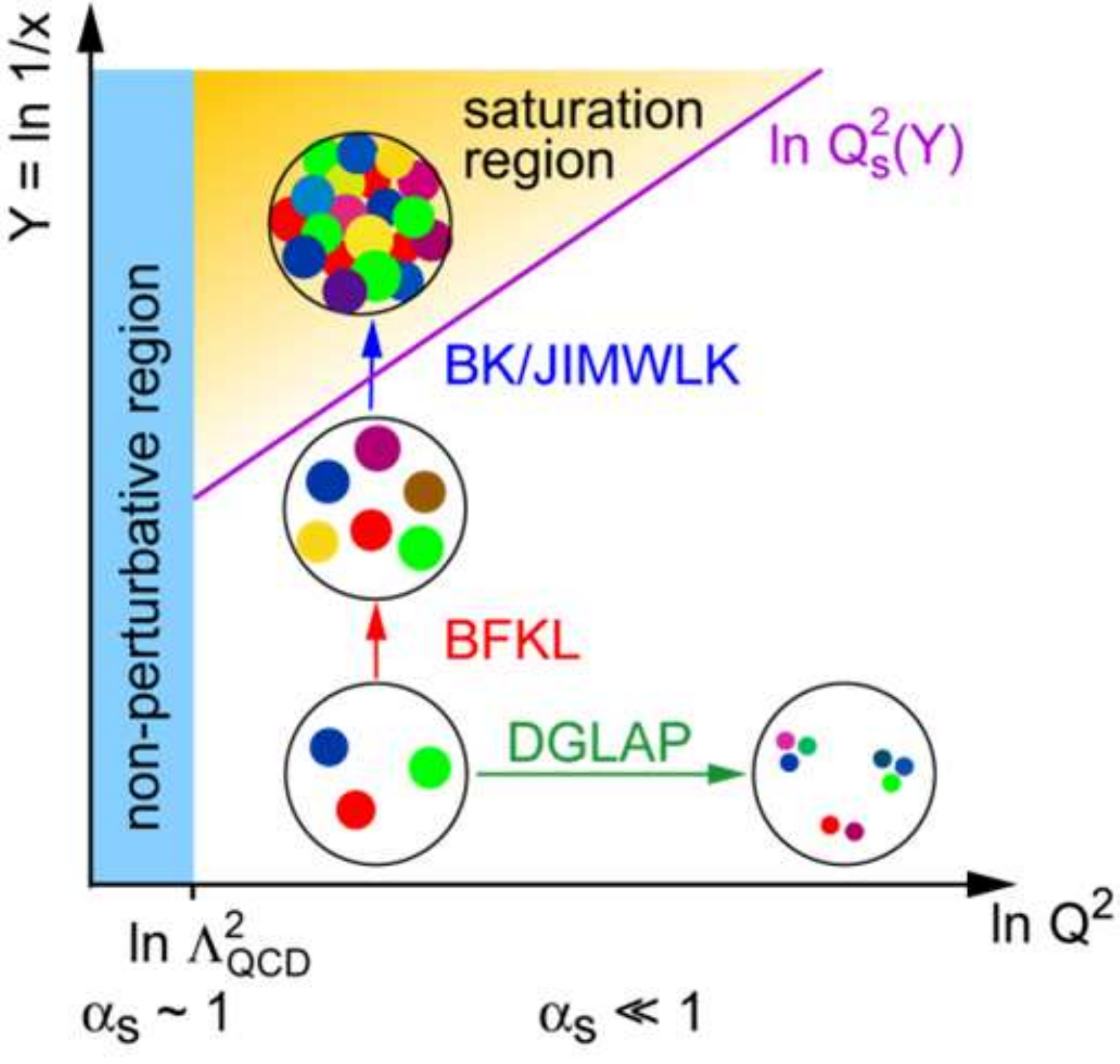}
  \caption{(left) Nuclear modification of gluons from a global analysis
    of deep-inelastic scattering (DIS), Drell-Yan, and PHENIX
hadron data by the EPS09 group~\cite{Eskola:2009uj}.
(right) Schematic diagram of the different regions in the momentum
    transfer ($Q^2$) vs momentum fraction ($x$) space, showing the
    gluon saturation region at large $\ln 1/x$ and small $\ln Q^2$.
}
  \label{fig:eps09}
\end{figure}

\begin{figure}[h]
\begin{minipage}{20pc}
\includegraphics[width=20pc]{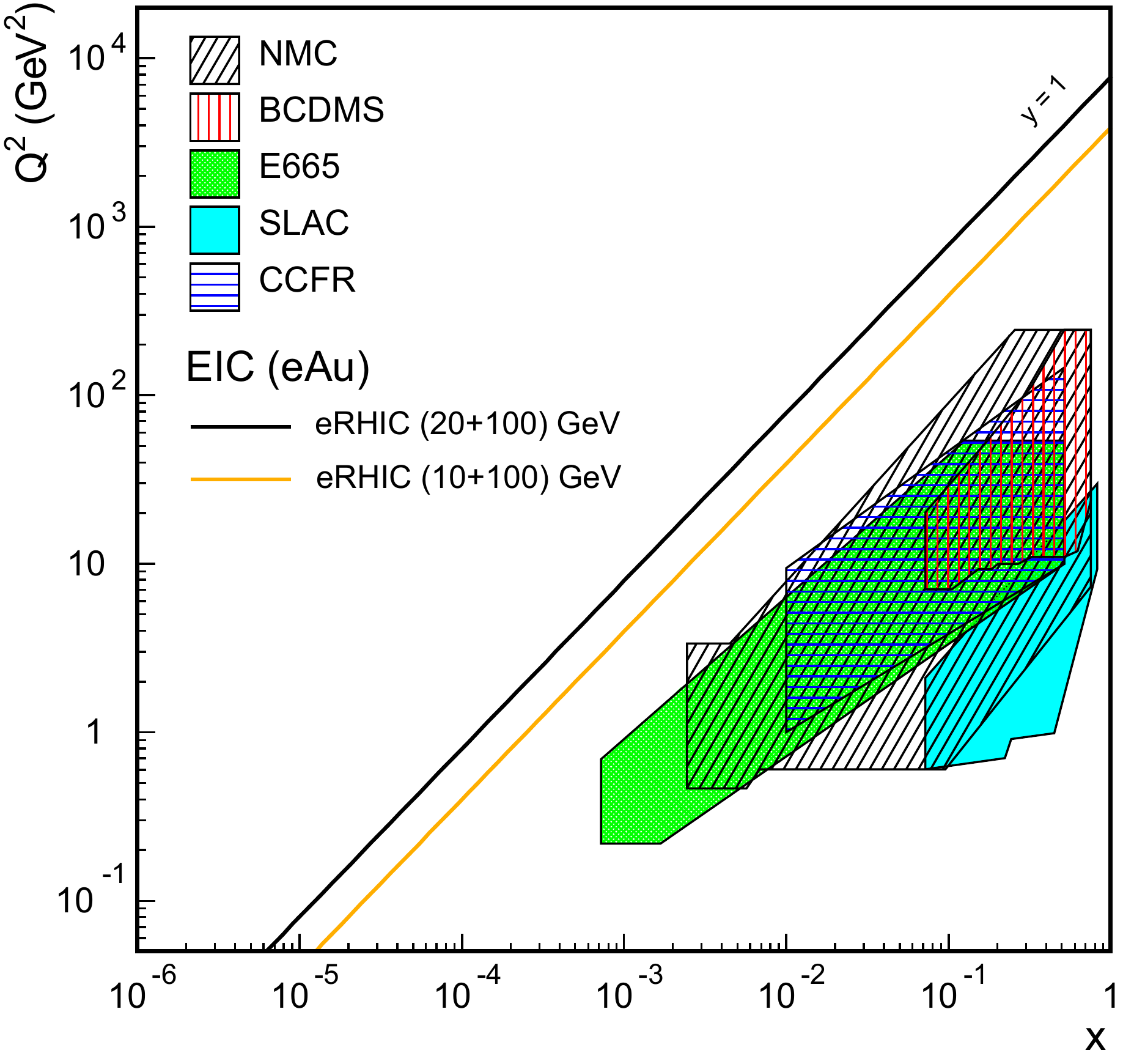}
\end{minipage}\hspace{2pc}%
\begin{minipage}{12pc}
\caption{\label{fig:xq2}The $x-Q^2$ coverage of electron-nucleus collisions at RHIC for two beam-energy combinations compared to different fixed-target experiments.}
\end{minipage}
\end{figure}

Electron-nucleus collisions at RHIC will further extend the existing battery of simpler collision systems regularly run at RHIC which aid in our understanding of complex heavy ion interactions.  As illustrated in the left panel of Fig.~\ref{fig:eps09}, the modification of the gluon distribution function in heavy nuclei is extremely uncertain, with depletion factors ranging from $\sim$10\% to nearly no gluons at $x \simeq 5 \times 10^{-3}$.  This uncertainty is due largely to the limited $Q^2$ range of earlier (fixed-target) nuclear deep-inelastic scattering (DIS) and Drell-Yan experiments, although inclusive pion production data from $d+$Au and $p+p$ collisions at RHIC are also included in recent global analyses of nuclear pdfs such as~\cite{Eskola:2009uj}.  High-energy, high-precision $e+A$ data for a range of nuclei would greatly reduce uncertainties on the initial-state gluon distribution, and further $d$ or $p+A$ data with multiple nuclear species would provide some sensitivity to the impact-parameter dependence of the nuclear pdfs as well as confirm and test our understanding of $e+A$ data.  The kinematic coverage in $x$ and $Q^2$ for lepton beam energies of 10 and 20 GeV at RHIC compared to previous nuclear DIS experiments is given in Fig.~\ref{fig:xq2}.

The problem of unitarity and the role of nonlinear evolution of parton distributions at
low $x$, generally referred to as gluon saturation, are among the key outstanding problems
in QCD.  Saturation models predict a regime where diagrams in which two low-momentum gluons fuse into one higher-momentum gluon become significant, thus depleting the low-momentum gluons. These effects are present in a nucleon, but become much stronger ($\propto A^{\frac{1}{3}}$) in a nucleus. Recent saturation models assert that a novel semiclassical state, the color glass condensate (CGC), is formed above a critical saturation scale ($Q_s^2$) at low enough momentum~\cite{Gelis:2010nm}. The right panel of Fig.~\ref{fig:eps09} shows a conceptual diagram of the CGC picture.  However, the $x$, $Q^2$ region where saturation physics plays a significant role is not yet well constrained.  In conjunction with continued $p(d)+A$, and $A+A$ collisions, electron-nucleus collisions could provide detailed information on a gluon saturation regime in nuclei.

\begin{figure}[h]
\begin{minipage}{15pc}
\includegraphics[width=15pc]{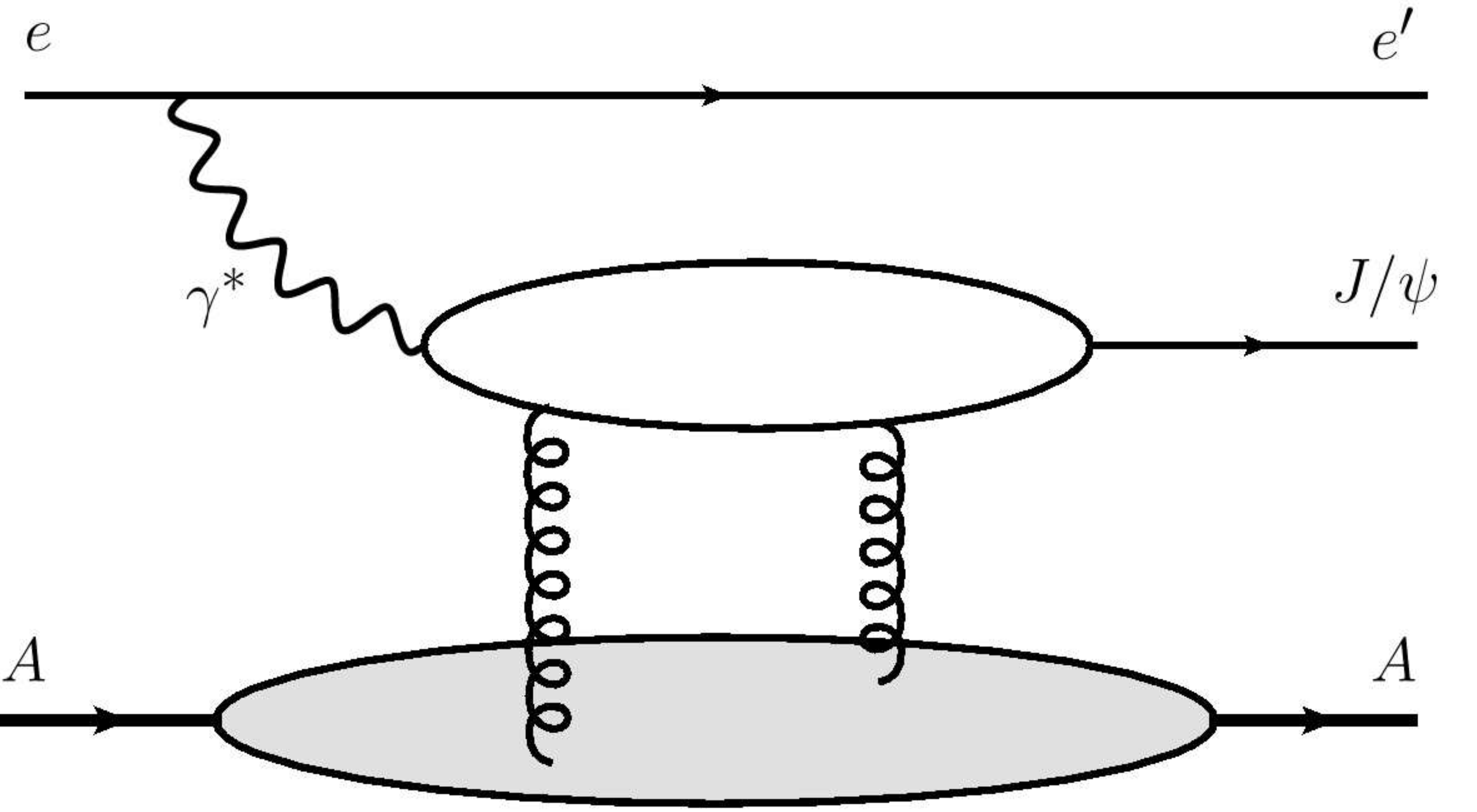}%
\end{minipage}\hspace{2pc}%
\begin{minipage}{15pc}
\caption{\label{fig:feynmanDiagram}Relevant Feynman diagram for diffractive vector meson production (here the $J/\psi$) in $e+A$. The $q\bar{q}$ dipole probes the gluon density in the nucleus $A$ with two gluons: not only can vector meson production probe the mean gluon density in the nucleus but also gluon density correlations.}
\end{minipage}
\end{figure}

\begin{figure}[h]
\centering
\begin{minipage}{14pc}
\includegraphics[width=14pc]{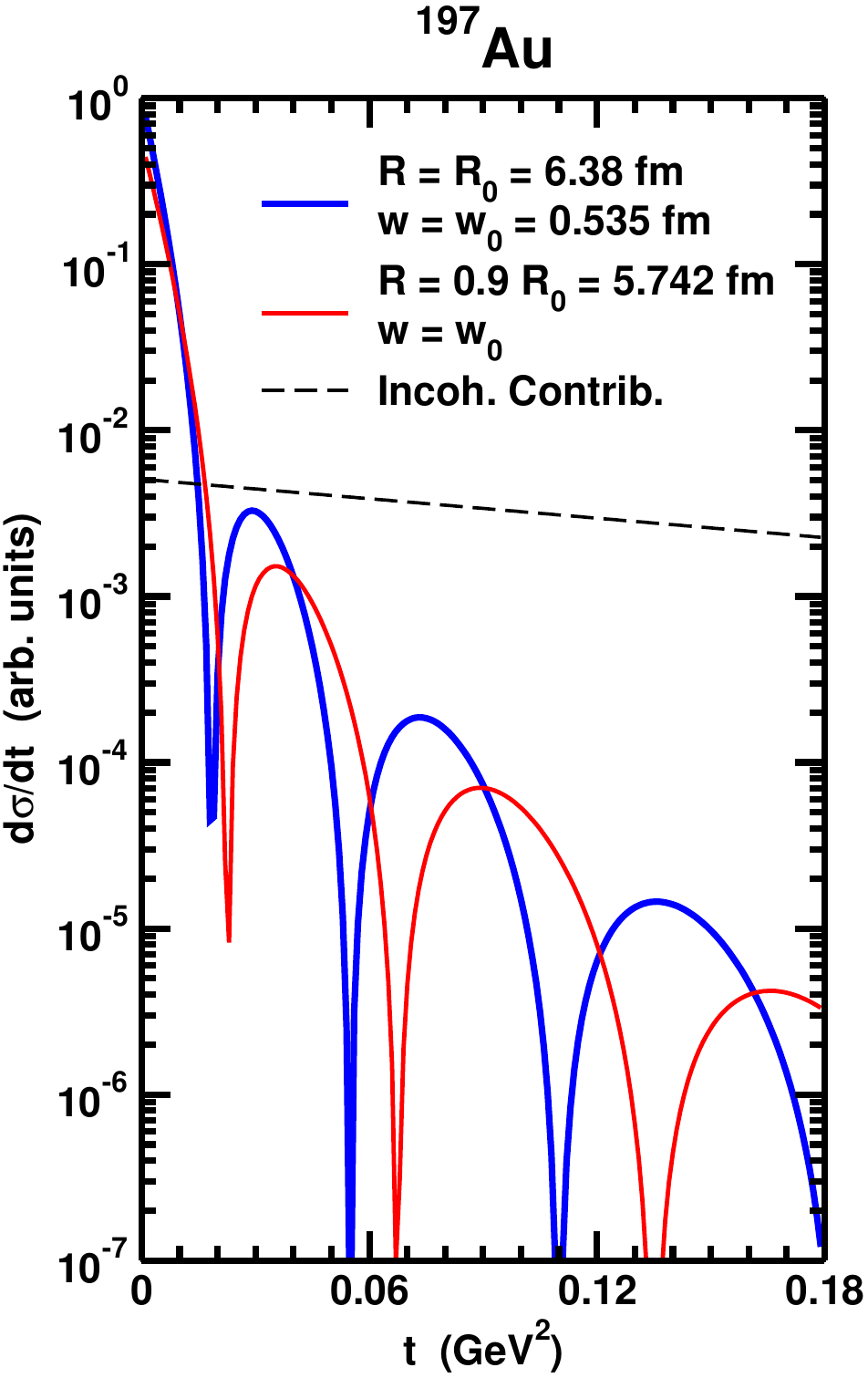}
\end{minipage}\hspace{4pc}%
\begin{minipage}{14pc}
\includegraphics[width=14pc]{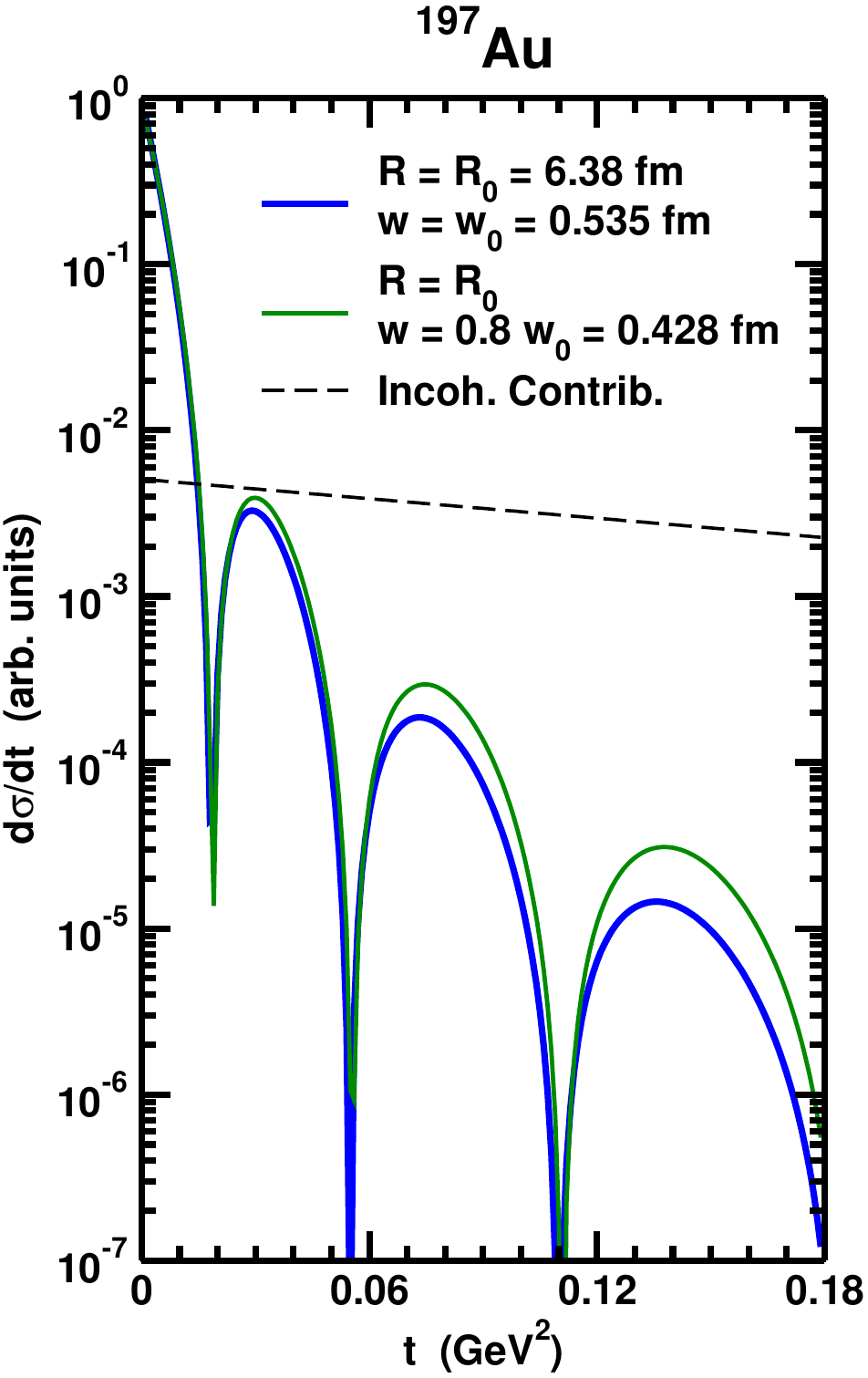}
\end{minipage}
\caption{\label{fig:diffraction}Vector meson diffraction patterns for $^{197}$Au. (blue curves) Differential cross section when the gluon density is assumed to scale with the Woods-Saxon charge distribution with the usual parameters taken for the radius and width. Shown are comparisons for a Woods-Saxon distribution with a (left) 10\% smaller radius and (right) 20\% smaller width. Both plots also show the dominant contribution to the cross section due to incoherent vector meson production. }
\end{figure}

Through diffractive measurements in electron-nuclear collisions, nuclear density distributions as a function of impact parameter can be studied.  Diffraction patterns measured in the processes of coherent vector meson production and deeply-virtual Compton scattering may be inverted to determine the initial nuclear gluon and quark densities.  The most sensitive density determination is at the edge of the nucleus, so these measurements will provide key information about the overlap region to which hydrodynamics and energy loss calculations are most sensitive.  The Feynman diagram contributing to coherent vector meson production, in which an electron emits a virtual photon that splits into a $q\bar{q}$ pair and subsequently interacts coherently with the nucleus to form the vector meson, is shown in Fig.~\ref{fig:feynmanDiagram}, and Fig.~\ref{fig:diffraction} shows vector meson diffraction patterns for gold.  The differential cross section for a gluon density assumed to scale with the Woods-Saxon charge distribution with the usual parameters taken for the radius and width is compared to the pattern for a Woods-Saxon distribution with a 10\% smaller radius and 20\% smaller width.  It can be seen that the measured coherent diffraction pattern is very sensitive to the details of the gluon density.  Also shown in the figure is the dominant contribution to the cross section coming from incoherent scattering, in which the outgoing nucleus is either broken up or in an excited state.

Proton-proton collisions at RHIC will remain important even after $e+p$ capabilities are in place.  The $p+p$ data will serve not only as a simpler system for comparison to heavy ion data, but also for continued study of proton structure, in particular its \emph{color} structure.  Study of the color structure of the nucleon and color interactions in hadronic collisions is a nascent field in QCD, for example with predictions of modified universality of certain (T-odd) $k_T$-unintegrated distribution functions due to different color interactions in semi-inclusive deep-inelastic lepton-nucleon scattering versus the Drell-Yan process~\cite{Collins:2002kn}.

\section{The sPHENIX Detector}
\begin{figure}[ht]
 \centering
  \includegraphics[width=1.0\linewidth]{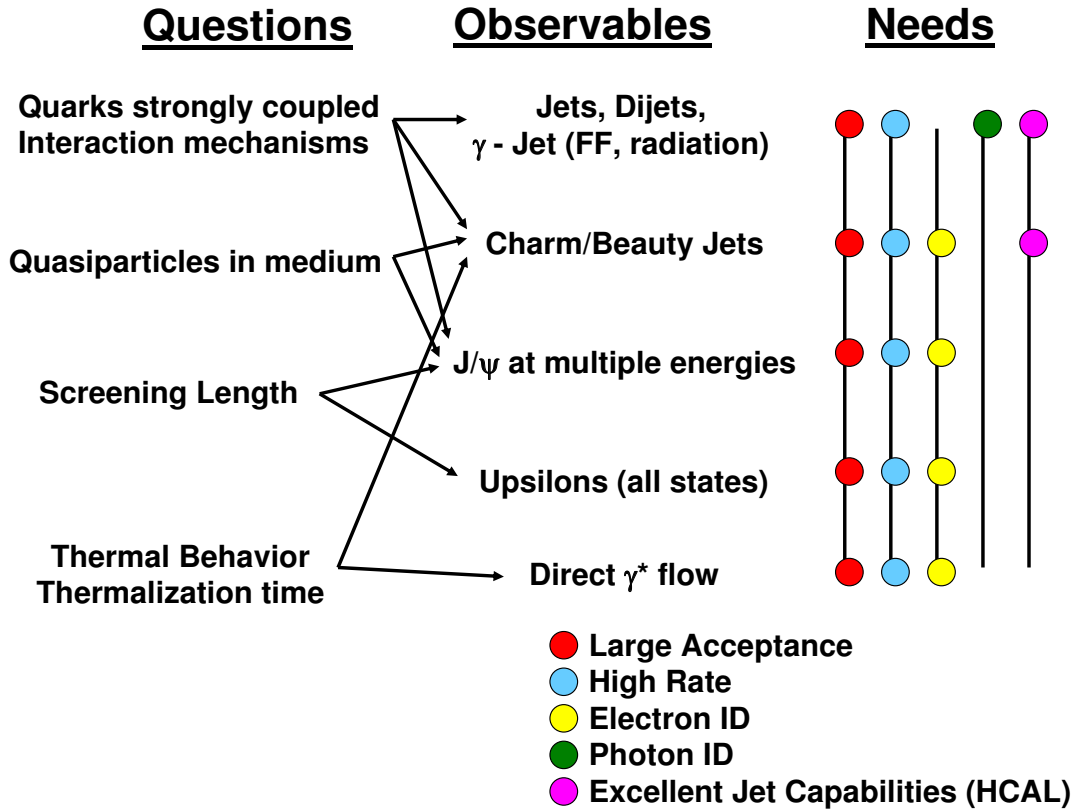}
  \caption{A chart illustrating the interesting physics questions after
    $\sim$2015, their relationship to final-state observables, and the
    detector requirements needed for these measurements.  }
  \label{fig:hifuturechart}
\end{figure}

The different physics questions of interest can be addressed by a variety of experimental observables, and these observables will determine the design of any upgraded detector at RHIC. Figure~\ref{fig:hifuturechart} illustrates heavy ion physics questions expected to remain open once the science program with currently planned or newly commissioned upgrades is completed around 2015.  It indicates their correspondence to a variety of observables and the associated detector requirements.

\begin{figure}[h]
\includegraphics[width=0.8\linewidth]{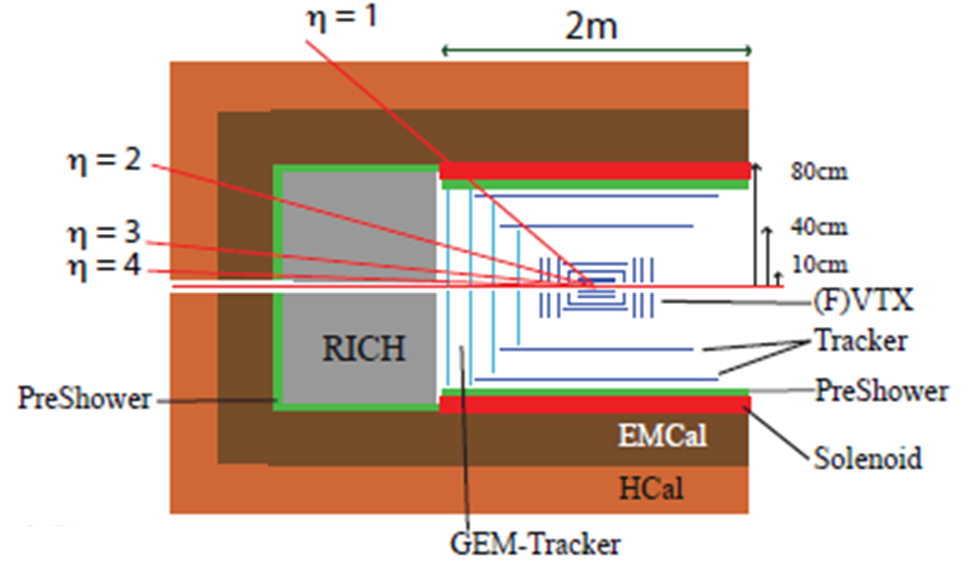}%
\caption{\label{fig:sPHENIX}A schematic layout of the sPHENIX detector.}
\end{figure}

\begin{figure}[h]
\begin{minipage}{14pc}
\includegraphics[width=14pc]{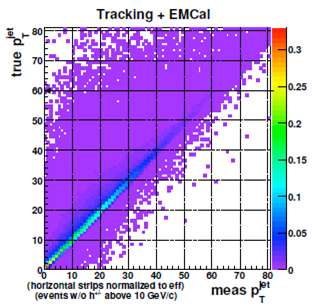}
\end{minipage}\hspace{2pc}%
\begin{minipage}{14pc}
\includegraphics[width=14pc]{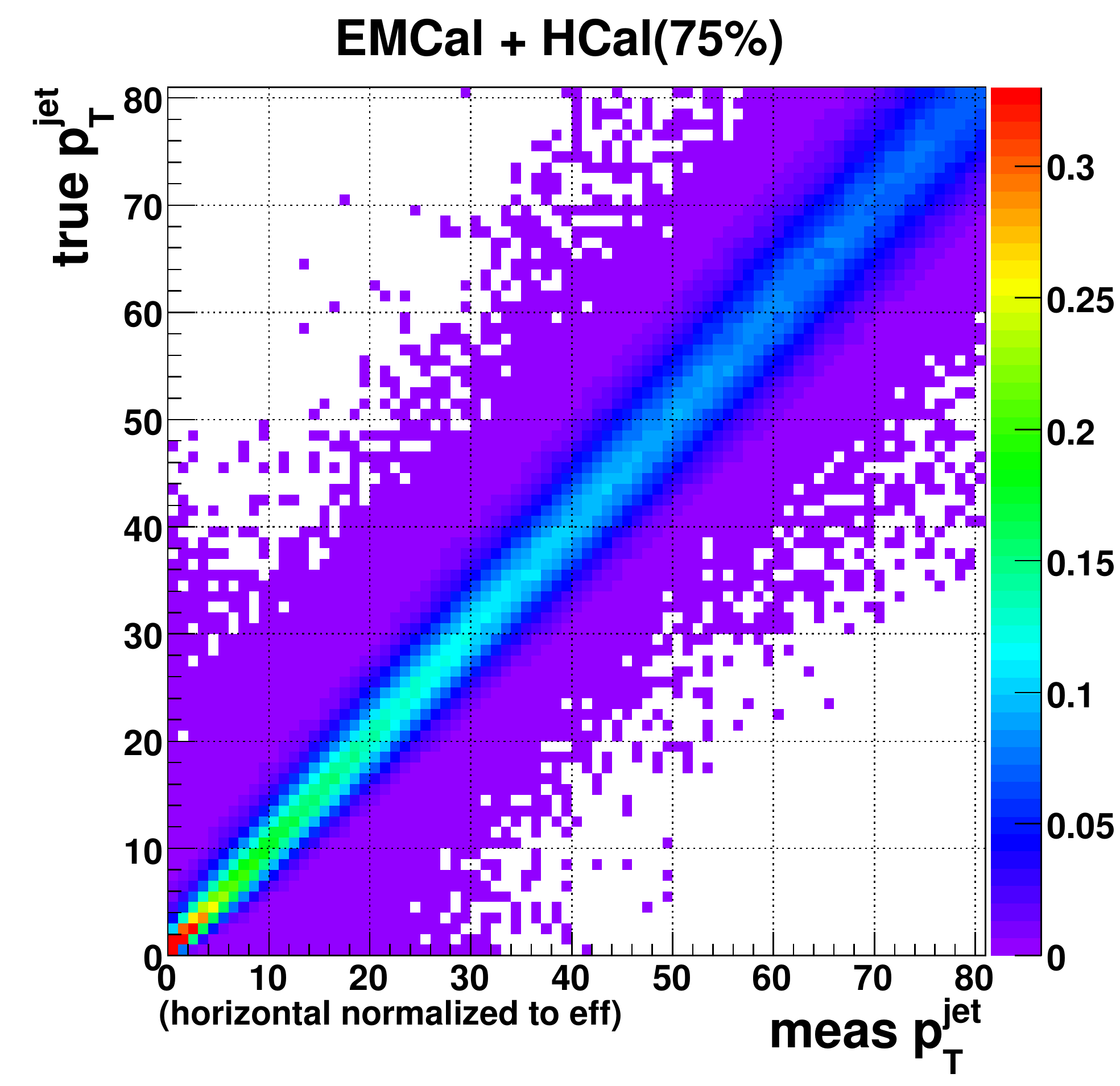}
\end{minipage}
\caption{\label{fig:jetHCal}Fast simulation response matrices demonstrating the improved correlation between measured and true $p_T$ of jets when tracking (left) is replaced by hadronic calorimetry with $\sigma_E/E = $75\% (right).
}
\end{figure}

Shown in Fig.~\ref{fig:sPHENIX} is a schematic design for an upgraded detector based on the requirements itemized in Fig.~\ref{fig:hifuturechart}.  The basic philosophy is to design a compact, uniform, nearly hermetic detector with fast rate capabilities and excellent full jet reconstruction at both mid- and forward rapidities.  The recently installed VTX silicon detector at midrapidity as well as the FVTX forward silicon detector, to be installed by the end of 2011, are retained.  Hadronic calorimetry at midrapidity provides improved jet energy resolution compared to tracking and electromagnetic calorimetry alone, and it increases the efficiency of jet reconstruction and triggering in order to extend measurements to high transverse momenta, as illustrated in Fig.~\ref{fig:jetHCal}.  A forward spectrometer covering up to $\eta \approx 4$ with tracking, particle identification, and both electromagnetic and hadronic calorimetry is envisioned.  Electromagnetic calorimetry up to large rapidities in the direction of the electron beam will be crucial to perform measurements in $e+p$ and $e+A$ collisions.  A large rapidity spectrometer will also be key for the study of asymmetric nuclear collisions as well as the investigation of many of the striking effects observed at high Feynman-$x$ related to parton dynamics in the nucleon.  One of the current PHENIX muon arms, not shown in Fig.~\ref{fig:sPHENIX}, would remain.

\section{Conclusion}

As the challenge of QCD continues, RHIC is well positioned to become an even more powerful tool to fulfill advancement to a quantitative era in QCD going into the 2020s.  As a dedicated QCD facility with capabilities for both electroweak as well as hadronic and nuclear collisions over a wide range of energies and system sizes, and with control over beam polarization for protons and possibly helium-3, the RHIC complex and its detectors would allow access to an impressive breadth of QCD physics.  Work is ongoing within the community to realize such a future.

\section*{References}
\bibliography{AidalaWWNDProceedings}

\end{document}